\documentclass[aps,prb,twocolumn,showpacs,superscriptaddress]{revtex4}
\usepackage{graphicx,amssymb,amsfonts,amsmath,bbm}

\begin{document}

\title{Current driven defect unbinding transition in an $XY$ ferromagnet}
\author{Aditi Mitra}
\affiliation{Department of Physics, New York University,
4 Washington Place, New York, New York 10003, USA}
\author{Andrew J. Millis}
\affiliation{Department of Physics, Columbia University, 538 W.
120th Street, New York, New York 10027, USA}
\date{\today}


\begin{abstract}
A Keldysh-contour effective field theory is derived  for magnetic vortices in the
presence of current flow.
The effect of adiabatic and non-adiabatic spin transfer torques on vortex motion is highlighted.
Similarities to and differences from the superconducting case are presented and explained.
Current flow across a magnetically ordered state is shown to lead to a defect-unbinding phase
transition
which is intrinsically nonequilibrium in the sense of not being driven by a variation in
effective temperature.
The dependence of the  density of vortices
on the current density is determined.
\end{abstract}


\maketitle

\section{Introduction}
The nonequilibrium physics of strongly interacting quantum systems is a topic of fundamental theoretical importance. A question of particular current interest  is the behavior of quantum critical systems under nonequilibrium conditions. Examples include phase transitions induced at temperature $T$=$0$ in the presence of a nonequilibrium  drive~\cite{Mitra06,Prosen08,Chung09,Oka10,Takei10,Torre10} as well as quantum coarsening and quench problems of a system prepared or maintained in a far-from equilibrium  state.~\cite{Aron09,Kirchner09,Gorczyca10,Polkovnikov10,Mitra11} In many of the cases of  nonequilibrium quantum criticality studied to date, the dominant physics is that the  nonequilibrium drive acts  to produce a noise term (typically delta-correlated at long scales)  in the equation of motion for the critical fluctuations.~\cite{Mitra06} This noise term increases fluctuations and destroys order, similarly to temperature.  Differences from equilibrium physics of course   appear, for example as violations of the fluctuation-dissipation theorem~\cite{Mitra05,Kirchner09} or as subleading corrections to scaling.~\cite{Mitra08,Takei10}  Other situations can occur too.  
For example, Dalle Torre {\it et al} showed that in some circumstances  departure from equilibrium can lead 
to a noise with long ranged 
correlations, which leads to larger differences from equilibrium physics.~\cite{Torre10}  In some models, excitations propagate ballistically so the concept of effective temperature  does not apply and nonequilibrium phase transitions arise for different reasons.~\cite{Feldman05,Prosen08,Gorczyca10}

In this paper we investigate an alternative route to  nonequilibrium quantum criticality. We show that in continuous-symmetry magnets an applied current can lead to a current-driven defect unbinding. This has no equilibrium analogue and therefore constitutes an intrinsically nonequilibrium phase transition, quantum in the sense that it is driven at $T$=$0$. Of course, even at $T$=$0$ a driving field will introduce a noise, but analogously to the nonequilibrium classical phase transitions occurring  e.g. in sheared liquids,~\cite{Onuki97} the noise is not the crucial parameter. The mechanism we discuss is generic, applying (with some difference of details) for any combination of  dimensionality and order parameter symmetry for which topological defects can be defined. For definiteness we present specific results in the case of a quasi two-dimensional $XY$ magnet, for which the defects involved are vortices. In this respect our work complements the treatment of current-driven quantum criticality  in our previous papers~\cite{Mitra06,Mitra08} which treated  Ising and Heisenberg symmetries and found that the transition was driven by a current-induced noise.

That an applied current can lead to a force on a magnetic defect was previously known.~\cite{Thiele73,Thiaville05,Shibata05} The transition we find is analogous to defect-unbinding transitions driven by supercurrent flow in superfluids.~\cite{Halperin80,Minnhagen87} The new contributions of this paper include a derivation from microscopics which is somewhat different from (although basically in agreement with) previous work,~\cite{Bazaliy98,Zhang04,Duine07,Nunez08,MacDonald09} a discussion of the mapping between the superconducting and magnetic situations, some additional insights into the relative importance of the different mechanisms by which Galilean invariance is broken and explicit results for the defect creation rates.

Our approach is to write a general interacting fermion model in the presence of a nonequilibrium drive as a Keldysh-contour path integral, introduce magnetic degrees of freedom via a Hubbard-Stratonovich transformation of the fermion-fermion interactions and by integrating out the fermions obtain a nonequilibrium spin model. We use a method presented by Schulz~\cite{Schulz88} to introduce rotational degrees of freedom and use an expansion in gradients and quantum fluctuations to obtain an action in terms of rotational spin excitations.The mean field approximation to this action is the Landau-Lifshitz-Gilbert equations. By representing  magnetic configurations in terms of topological (vortex) coordinates~\cite{Shibata05} we obtain an action for the topological excitations, which is then analyzed via quantum critical and instanton methods.

The rest of this paper is organized as follows. Section~\ref{Formalism} outlines the approach we use,
Sections~\ref{Drive} and~\ref{Noise} present a derivation of the basic dynamical effects, Section~\ref{Action} summarizes the resulting action  in spin  coordinates, Section~\ref{Topological} derives the action for topological defects, Section~\ref{Transition} describes the current-driven vortex unbinding transition and Section~\ref{Conclusion} is a summary and conclusion.

\section{Formalism \label{Formalism}}

\subsection{Overview}

We are interested in the effect of current drive on the nucleation and  motion of topological defects in the magnetization field of a metallic ferromagnet.  Although our formalism can be applied to the case of any topological defect, we present explict results for vortices: the topological defects in a two dimensional easy-plane magnet. Because topological defects are large scale structures in the spin field, they obey equations of motion which may be deduced from semiclassical arguments based on the Landau-Lifshitz-Gilbert equation.~\cite{Thiele73,Zhang04,Thiaville05} 
Denoting the position of defect $i$ by ${\vec X}_i$ and the sign of  its vorticity by $\eta=\pm1$, we may represent its equation of motion as

\begin{equation}
\eta {\cal C}{\cal J}_0 {\hat z} \times {\dot {\vec X}}_i+I_D {\cal D}_0\dot {\vec X}_i=\eta {\cal C} {\hat z} \times {\vec {\cal J}}+I_D{\vec {\cal D}}+\sum_j{\vec F}_{ij}+{\vec \zeta}_i
\label{vortexEOM1}
\end{equation}
with   ${\vec F}_{ij}$ the force due to the intervortex interaction (including both spin-wave-mediated and dipolar terms). The terms ${\vec{\mathcal J}}$ and ${\vec {\mathcal D}}$ arise due to current flow, while ${\cal J}_0$ and ${\cal D}_0$ representing the effects of spin conserving and non-conserving processes on the temporal dynamics. $I_D$ and ${\mathcal C}$ are constants and $\zeta_i$ is a fluctuating field which on scales longer than the mean free path may be taken to be delta correlated, with
\begin{equation}
\left< \zeta^a_i (t)\zeta^b_j(t{'}\right>=2I_DN^{xx}T^*\delta_{ij}\delta^{ab}\delta(t-t{'})
\label{noisecorr1}
\end{equation}
Here  $N^{xx}$ is related to the many-body density of states and $T^*$ is an effective temperature  arising from the departures from equilibrium.

In this paper we employ a Keldysh-contour based nonequilibrium approach to calculate the coefficients in Eq.~(\ref{vortexEOM1}) in terms of the underlying properties of the electrons (in particular the transport and spin relaxation times) and the applied current, determining the relative magnitudes of the different terms and explicating the similarities and differences between topological defects in superconductors and in magnets. 

The formalism is based on a phase angle representation convenient for carrying out a gradient expansion. It enables a quantitative understanding of the origin of the breaking of Galilean invariance, of the roles of spin conserving and non-conserving scattering mechanisms, and of the variations of parameters as the magnetic critical point is approached. We show that in the generic situation when the spin non-conserving  scattering is small compared to the spin conserving scattering, then  the vortex Hall angle is typically near to $90^\circ$ (in contrast to the result for superconductors), but the drag contribution to the force on a vortex is much larger than the transverse (`Magnus') force contribution (also different from the usual case of vortices in superconductors).  These results, in combination with the effects of the long-ranged dipole interaction, lead to quantitative differences in the current-induced nucleation of vortices.

\subsection{Basic Assumptions}

We consider  a system of interacting fermions and make the Fermi liquid assumption that the dynamical  effects of the electron-electron  interaction may be neglected at the scales of interest here. The system is then characterized by a band structure, which we represent via a hopping matrix $t_{i-j}$ connecting unit cells $i$ and $j$,  and a local magnetization
\begin{equation}
{\vec m}(r,t)={\hat {\mathbf n}}({\vec r},t)m({\vec r},t)
\label{mdef}
\end{equation}
of direction ${\hat {\mathbf n}}$ and magnitude $m$.

The physics also involves a self energy ${\mathbf \Sigma}$ which encodes the effects of scattering processes.   We will distinguish two sources of scattering: spin conserving processes corresponding e.g. to scattering off of spin-independent random potentials and spin non-conserving processes:
\begin{equation}
{\mathbf \Sigma }_{tot}={\mathbf \Sigma}_{cons}+{\mathbf \Sigma}_{N-C}
\label{sigmasplit}
\end{equation}
We will assume that the spin non-conserving processes arise from electron exchange with a spin-unpolarized reservoir which does not change when a current is applied to the system.~\cite{Mitra08} Thus ${\mathbf \Sigma}_{N-C}$ is proportional to the unit matrix in spin space and is independent of applied current.  Because for example the density of states may be spin dependent ${\mathbf \Sigma}_{cons}$ may have structure in spin space. For simplicity we will assume that the magnetization varies sufficiently slowly in space that the electrons always relax to the local magnetization; in this case ${\mathbf \Sigma}_{cons}$ is the sum of a term proportional to the unit matrix and a term proportional to ${\vec m}\cdot {\vec \sigma}$ with, in particular, no terms proportional to gradients of ${\vec m}$. Where an explicit model is necessary we will take  ${\mathbf \Sigma}_{cons}$ to arise from randomly placed point-like impurities.

We suppose that the system is driven out of equilibrium by an externally imposed current which is weak enough to be described by linear response theory.  Since we are interested in length scales long compared to the spin relaxation length, any spin polarized current injected at the boundary will quickly relax.  The relevant currents are thus induced by applying an electric field, ${\vec E}$. In the presence of a non-vanishing magnetization the resistivity ${\mathbf \rho}$ will be a matrix in spin space so the  current density ${\vec J}=\mathbf{\rho}^{-1}{\vec E}$ will have both a charge and a spin component. Our assumption that electrons are relaxed to the local magnetization direction means that the spin direction is aligned to the local magnetization so that  in the basis aligned with the local magnetization ${\mathbf \rho}$ has components $\rho_\sigma$ which may differ for up and down spins and (denoting the Cartesian component of a vector by a roman symbol)
\begin{equation}
{\mathbf J}^a=J_0^a{\mathbf 1}+J_M^a{\hat{\mathbf n}}\cdot{\vec \sigma}
\label{Jfinal}
\end{equation}

In the Keldysh two contour Greens function method the current flowing between sites $i$ and $j$  is determined by the Keldysh component ${\mathbf G}^K$ of the electron Greens function via
\begin{equation}
{\mathbf J}_{ij}^a= -i{\mathbf G}^K(i,j;t,t){J}_{ji}^a
\label{spincurrent}
\end{equation}
with current operator
\begin{equation}
{\vec J}_{i-j}=-it_{i-j}\left({\vec r}_i-{\vec r}_j\right)
\label{Jdef}
\end{equation}
Here $\left({\vec r}_i-{\vec r}_j\right)$ is the vector connecting sites $i$ and $j$ and in momentum space ${\vec J}=\partial \varepsilon_k/\partial {\vec k}\equiv {\vec v}_k$

The retarded and advanced components of the Greens function and self energy  change only to second order in departures from equilibrium, so within the linear response approximation, departures from equilibrium may be described by nonequilibrium terms in the  Keldysh components of the electron Greens function ${\mathbf G}$ and self energy ${\mathbf {\Sigma}}$. Thus we have e.g.  ${\mathbf G}^K\rightarrow {\mathbf G}^K_{eq}+{\mathbf G}^K_{neq}$, with ${\mathbf G}^K_{neq}$ calculated to linear order in the applied field. The nonequilibrium terms are discussed in detail in section~\ref{Noise}; here we note simply that in a simple electronically three dimensional  model with $k^2/2m$ dispersion, magnetization aligned along the spin-$z$ axis and relatively weak scattering, the formalism leads to the familiar Drude formula
\begin{equation}
\rho_\sigma=\frac{m \Gamma^{tot}_\sigma}{n_\sigma}
\label{rhosimple}
\end{equation}
with $\Gamma^{tot}_\sigma$ the possibly spin-dependent scattering rate obtained from $Im{\mathbf \Sigma}_{tot}(\omega=0)$ and $n_\sigma$ the total number of electrons with spin aligned (or anti-aligned) to the local magnetic field.

Also, as discussed in Ref.~\onlinecite{Mitra08} an applied electric field leads to a noise term corresponding to an effective temperature
\begin{equation}
T_{eff}=eEl^*
\label{Teffdef}
\end{equation}
related to the voltage dropped over an appropriate length scale $l^*$. The differences in physics between the model considered here and the model studied in Ref.~\onlinecite{Mitra08} will lead to a different expression for $l^*$.

\subsection{Nonequilibrium terms in Greens function  \label{Distribution}}

In this subsection we sketch a derivation of the nonequilibrium terms in the electron Greens function and self energy. The treatment follows Ref.~\onlinecite{Mitra08}. Noting that the retarded and advance self energies are unchanged  to linear order in applied fields, we parametrize the change in the Keldysh components of ${\mathbf G},{\mathbf \Sigma}$ via a distribution function $h$ related to the Greens function  via ${\mathbf G}^K={\mathbf G}^Rh-h{\mathbf G}^A$ and similarly for ${\mathbf \Sigma}$. We then follow the usual steps to obtain  a quantum kinetic equation for $h$. In writing the kinetic equation we make use of our assumption that the electron distribution relaxes to the local magnetization so that we have one kinetic equation for each spin direction.  We further observe that  in the model considered here,  impurity scattering may relax the momentum dependence of the distribution function but not the energy dependence.  We write $h=h_{eq}+h^\sigma_A+h^\sigma_S$ with $h_{eq}(x)=1-2\Theta(-x)$ and nonequilibrium parts  $h_A$ odd and $h_S$ even  under $k \leftrightarrow -k$. Separating the even and odd parts of the kinetic equation we obtain (not denoting the spin direction explicitly)
\begin{equation}
\frac{\partial h_S}{\partial x}\left({\vec E}\cdot {\vec v}\right)+2\Gamma_{tot}h_A=
-2\delta(x)\left({\vec E}\cdot {\vec v}\right)
\label{odd}
\end{equation}
and
\begin{equation}
\frac{\partial h_A}{\partial x}{\vec E}\cdot {\vec v}+2\Gamma_{N-C}h_S=0
\label{even}
\end{equation}

Solving  and restoring spin indices explicitly we find
\begin{equation}
h_A^\sigma=-\frac{{\vec E}\cdot {\vec v}}{\Gamma^\sigma_{tot}}\left(\frac{\sqrt{\Gamma_{N-C}
\Gamma^\sigma_{tot}}}
{\left|{\vec E}\cdot {\vec v}\right|}Exp\left[-\frac{2\left|x\right|
\sqrt{\Gamma_{N-C}\Gamma_{tot}^{\sigma}}}{\left|{\vec E}\cdot {\vec v}\right|}\right]\right)
\label{hA}
\end{equation}
and
\begin{equation}
h_S^\sigma=-2\frac{sgn(x)}{2}
Exp\left[-\frac{2\left|x\right|\sqrt{\Gamma_{N-C}\Gamma^\sigma_{tot}}}{\left|{\vec E}\cdot {\vec v}\right|}\right]
\label{hS}
\end{equation}
These expressions differ from those obtained in Ref.~\onlinecite{Mitra08} because in that work $\Gamma^{tot}=\Gamma_{N-C}$.

For small $E$ the term in parentheses in Eq.~(\ref{hA}) is just an integral representation of the delta function $\delta(x)$. To an accuracy sufficient for the leading terms in the semiclassical expansion (Sec.~\ref{Drive}),  Eq.~(\ref{hA}) implies
\begin{equation}
{\mathbf G}^K_{neq}(i,l,\omega)=2i{\mathbf G}^R(i-j,0){\vec J}_{j-k}\cdot {\vec E}\delta(\omega){\mathbf G}^A(k-l,0)
\label{GKneqfinal}
\end{equation}
and
\begin{equation}
{\mathbf \Sigma}^K_{neq}=2i{\vec J}_{j-k}\cdot {\vec E}\delta(\omega)\label{SigmaneqKfinal}
\end{equation}

For small $E$, $h_S$ is just two powers of $E$ times an integral representation of the derivative of the delta function so terms involving $h_S$ would seem to give only contributions of higher order in $E$. But as shown in Ref.~\onlinecite{Mitra08} and discussed in more detail below,  the singularity in the $T$=$0$ equilibrium distribution function means that the $h_S$ terms give a contribution of order $|E|$ to the noise. This issue will be further discussed in Section~\ref{Noise}.

\subsection{Path Integral Representation}
We write the model as a path integral on the Keldysh two-time contour.~\cite{Kamenevrev,Mitra06} For conducting systems which have magnetic ground states we expect that part of the interaction term may be decoupled by a Hubbard-Stratonovich transformation~\cite{Schulz88} while the remainder can be treated by Fermi liquid theory. The important part of the Hubbard-Stratonovich field is a vector in spin space ${\vec m}(r,t)\cdot {\vec \sigma}$ which represents the local magnetization.

After integrating out the fermions we may express the model as a path integral
\begin{equation}
Z=\int {\mathcal D}{\vec m}e^{iS_F[\{m(r,t)\}]+iS_{int}[\{m(r,t)\}]}
\label{Zdef}
\end{equation}
In Eq.~(\ref{Zdef}) we consider time as running along the full Keldysh contour (including forward and  backward-moving parts and with 'Keldysh' time ordering assumed). The term $S_{int}[\{{\vec m}(r,t)\}]$ includes other terms in the energetics of the local spin polarization, including the terms arising from the Hubbard-Stratonovich transformation and the magnetic dipole energy. $S_F$, the logarithm of the fermion determinant, is
\begin{equation}
S_F=-iTr \ln \left[{\mathbf G}_0^{-1}-m{\hat {\mathbf n}}_i(t)\cdot {\vec {\bf \sigma}}\tau_3
-\tau_3{\mathbf \Sigma}_{tot}\tau_3\right]
\label{SFdef}
\end{equation}
$S_F$ depends on ${\mathbf G}_0^{-1}$, which  is a matrix in space, time and Keldysh contour indices and expresses  the effects of band structure and of  Fermi liquid renormalizations. The matrix $\vec{\tau}$ acts in Keldysh space.

Following Schulz~\cite{Schulz88} we now rotate the spin quantization axis  at each space-Keldysh time point so that it is parallel
to ${\mathbf {\hat n}}$.  The set of space-time rotation matrices which do this are ${\mathbf R}_{{\hat {\mathbf n}}}$ defined so that
${\mathbf R}^\dagger_{{\hat {\mathbf n}}}{\hat {\mathbf n}}\cdot {\vec \sigma} {\mathbf R}_{{\hat {\mathbf n}}}=\sigma_z$. Writing the spin orientation vector ${\hat n}({\vec r},t)=cos\theta {\hat z}+sin\theta { \hat u}$ with ${\hat  u}$ a vector in the $x-y$ plane, it is convenient to represent ${\mathbf R}$ in terms of  ${\hat \Omega}$ defined by
\begin{equation}
{\hat  \Omega}=\left(cos\frac{\theta}{2} {\hat {\mathbf z}} +sin \frac{\theta}{2} {\hat {\mathbf u}}\right)
 \label{omegadef}
 \end{equation}
 as
 \begin{equation} {\mathbf R}={\hat \Omega}_z{\mathbf 1}+i{\hat z}\times {\hat \Omega}\cdot {\mathbf \sigma}
 \end{equation}
We obtain
\begin{equation}
S_F=-iTr \ln \left[{\mathbf R}^\dagger_{{\hat {\mathbf n}}}
{\mathbf G}_0^{-1}{\mathbf R}_{{\hat {\mathbf n}}}-m{\mathbf \sigma}_z
-{\mathbf R}^\dagger_{{\hat {\mathbf n}}}\tau_3{\mathbf \Sigma}_{tot}\tau_3
{\mathbf R}_{{\hat {\mathbf n}}}\right]
\label{Srotated}
\end{equation}

The spin rotations correspond to an $SU(2)$ gauge transformation of the theory. If all of the terms in the underlying Hamiltonian  are  invariant under rotations of spin in the system of interest then the trace must be independent of the choice of spin quantization axis, corresponding to a gauge invariance of the model. Terms such as the local magnetization $m{\mathbf {\hat n}}\cdot {\vec {\mathbf \sigma}}$ or  sources (not explicitly written here) will break the $SU(2)$ invariance and lead to dependence on the spin directions. The reservoir which provides spin relaxation also breaks the $SU(2)$ gauge invariance because as we have defined it the rotation does not act on these degrees of freedom. Thus for example a time dependent spin rotation corresponds to a spin-dependent phase difference between system and reservoir, which could drive spin currents between the system of interest and the reservoir. A similar situation obtains in the theory of dissipative effects in superconductors.~\cite{Halperin80,Mitra08b}

From Eq.~(\ref{Srotated}), using ${\mathbf R}^\dagger_{{\hat {\mathbf n}}}{\mathbf R}_{{\hat {\mathbf n}}}=1$, and defining the rotated frame Greens function ${\mathbf G}$ via ${\mathbf G}^{-1}=\left({\mathbf G}_0^{-1}-m\sigma_z-\tau_3{\mathbf \Sigma}\tau_3\right)$ we obtain
 \begin{equation}
 \label{Srot}S_F=-iTr \ln \left[{\mathbf G}^{-1}+\delta{\mathbf  H}-\delta{\mathbf\Sigma}\right]
 \end{equation}
Here
\begin{equation}
\delta{\mathbf  H}={\mathbf  R}_{{\hat {\mathbf n}}}^\dagger{\mathbf G}_0^{-1}{\mathbf R}_{{\hat {\mathbf n}}}
-{\mathbf G}_0^{-1}
\label{deltaHeq}
\end{equation}
and
\begin{equation}
\delta{\mathbf \Sigma}={\mathbf R}^\dagger_{{\hat {\mathbf n}}}\tau_3{\mathbf
\Sigma}\tau_3{\mathbf R}_{{\hat {\mathbf n}}}-\tau_3{\mathbf \Sigma}\tau_3
\label{deltaSigma}
\end{equation}
Assuming the magnetization varies slowly in space and expanding to second order in rotations gives
\begin{equation}
S=S_0+S_1+S_2
 \label{Sexpand}
\end{equation}
with
\begin{eqnarray}
S_0&=&-iTr ln \left[{\mathbf G}^{-1}\right]+S_{int}
\label{S0def}
\\
S_1&=&-iTr\left[{\mathbf G}\left(\delta{\mathbf H}-\delta {\mathbf \Sigma}\right)\right]
\label{S1def}
\\
S_2&=&\frac{i}{2}Tr\left[{\mathbf G}\left(\delta{\mathbf H}-\delta {\mathbf \Sigma}\right){\mathbf G}\left(\delta{\mathbf H}-\delta {\mathbf \Sigma}\right)\right]
\label{S2def}
\end{eqnarray}

At this point we write ${\mathbf G},{\mathbf \Sigma}\rightarrow {\mathbf G}_{eq}+{\mathbf G}_{neq},{\mathbf\Sigma}_{eq}+{\mathbf \Sigma}_{neq}$ and retain terms up to first order in departures from equilibrium. Considering first the equilibrium terms we observe that our assumptions  of non-magnetic scattering, slow variation of the magnetization and relaxation of quasiparticle distribution to the local magnetization imply that only the terms ${\mathbf \Sigma}_{N-C}$ arising from spin non-conserving quantities contribute to $\delta{ \mathbf \Sigma}_{eq}$:
\begin{equation}
\delta{ \mathbf \Sigma}_{eq}=\delta{ \mathbf \Sigma}^{eq}_{N-C}={\mathbf R}^\dagger_{{\hat {\mathbf n}}}
\tau_3{\mathbf \Sigma}^{eq}_{N-C}\tau_3{\mathbf R}_{{\hat {\mathbf n}}}-\tau_3{\mathbf \Sigma}^{eq}_{N-C}\tau_3
\end{equation}
Thus for the equilibrium terms we have
\begin{equation}
\delta S^{eq}=S_1^{eq}+S_2^{eq}
 \label{Seqdecompose}
\end{equation}
with
\begin{eqnarray}
S_1^{eq}&=&-iTr\left[{\mathbf G}_{eq}\left(\delta{\mathbf H}-\delta {\mathbf \Sigma}_{N-C}^{eq}\right)\right]
\label{S1eqdef}
 \\
S_2^{eq}&=&\frac{i}{2}Tr\bigg[{\mathbf G}_{eq}\left(\delta{\mathbf H}-\delta {\mathbf \Sigma}_{N-C}^{eq}\right)
\nonumber
\\
&&\hspace{.3in}\times{\mathbf G}_{eq}\left(\delta{\mathbf H}-\delta {\mathbf \Sigma}_{N-C}^{eq}\right)\bigg]
\label{S2eqdef}
\end{eqnarray}
Here $S_1^{eq}$ will be seen to give the basic Berry phase magnetization dynamics while $S_2^{eq}$ gives the spin stiffness.

Turning now to the nonequilibrium terms, we work to leading order in both departures from equilibrium and rotations, neglecting the unimportant change in spin stiffness due to current flow. We obtain
\begin{equation}
S_1^{neq}=-iTr\left[{\mathbf G}_{neq}\left(\delta{\mathbf H}-\delta {\mathbf \Sigma}_{N-C}^{eq}\right)\right]
+iTr\left[{\mathbf G}_{eq}\delta {\mathbf \Sigma}_{neq}\right]
\label{Sneqdef}
\end{equation}

The rest of this paper is based on an  evaluation of 
Eqs.~(\ref{S0def}),~(\ref{S1eqdef}),~(\ref{S2eqdef}),~(\ref{Sneqdef}) in the presence of spin textures and externally imposed currents and in the limit of slow variation of spin direction. We perform an expansion in the amplitude of the quantum components of the rotation matrices. The leading terms in the semiclassical expansion give rise to the classical equations of motion for spin fluctuations and topological defects in the presence of an applied current. The next-to-leading terms lead to a current-induced effective temperature.

\vspace{0.1in}

\section{Drive and Dynamics\label{Drive}}

\subsection{Overview}
In this section we present the leading terms in a semiclassical expansion.  These terms give rise to the classical equations of motion for spin fluctuations and topological defects in the presence of an applied current. They
are most easily obtained by going to the  Larkin basis in contour space,~\cite{Kamenevrev} introducing classical (C) and quantum (Q) combinations of the time-ordered ${\mathbf R}_2$ and anti-time ordered ${\mathbf R}_1$ Keldysh contour fields via
\begin{eqnarray}
{\mathbf R}_C&=&\frac{1}{2}\left({\mathbf R}_2+{\mathbf R}_1\right)
\\
{\mathbf R}_Q&=&\frac{1}{2}\left({\mathbf R}_2-{\mathbf R}_1\right)
\end{eqnarray}
and expanding to leading order in ${\mathbf R}_Q$. We evaluate the terms in order.

\subsection{$S_1^{eq}$}
We first consider the  terms involving ${\mathbf \delta H}$. Expanding and taking the trace over contour indices gives
\begin{eqnarray}
S_{1}^{eq,G_0}&=&-iTr\bigg[{\mathbf G}^K(1-2){\mathbf R}^\dagger_C(2)G_0^{A,-1}(2-1){\mathbf R}_Q(1)
\nonumber \\
&+&{\mathbf G}^K(1-2){\mathbf R}^\dagger_Q(2)G_0^{R,-1}(2-1){\mathbf R}_C(1)\bigg]
\label{S1eq0}
\end{eqnarray}
where the trace is over space, time and spin degrees of freedom.

Now $G_0^{R,-1}({\vec r}_1-{\vec r}_2;t_1-t_2)=\delta(t_1-t_2)\left(i{\overrightarrow \partial_t}-t_{ij}\right)$ and  $G_0^{A,-1}({\vec r}_1-{\vec r}_2;t_1-t_2)=\left(-i{\overleftarrow \partial_t}-t_{ij}\right) \delta(t_1-t_2)$ are short ranged in space and time so we may perform a gradient expansion of Eq.~(\ref{S1eq0}). Because ${\mathbf R}^\dagger_Q(1){\mathbf R}_C(1)+{\mathbf R}^\dagger_C(1){\mathbf R}_Q(1)= ({\mathbf R}^\dagger_2(1){\mathbf R}_2(1)-{\mathbf R}^\dagger_1(1){\mathbf R}_1(1))/2$ vanishes, the leading terms are the first derivatives. The space derivative term vanishes because in equilibrium $G^K$ is an even function of its space arguments. Thus we obtain
\begin{equation}
S_{1}^{eq,G0}=-iTr\left[{\mathbf G}^K_0\left(\bigl\{-i\partial_t{\mathbf R}^\dagger_C\bigr\}
{\mathbf R}_Q+{\mathbf R}^\dagger_Q\bigl\{i\partial_t{\mathbf R}_C\bigr\}\right)\right]
\label{SG0}
\end{equation}
$-i{\mathbf G}_0^K\equiv -i{\mathbf G}^K(i,i;t,t)=\rho+m\sigma^z$ with $\rho$ a scalar related to the charge density and $m$ the magnitude of the magnetization on site $i$ at time $t$. ${\hat z}$ is the (arbitrarily chosen) spin quantization axis.

We next consider the term in $S_1^{eq}$ arising from the equilibrium self energy. Expressing the result in the Larkin basis, taking the trace and  retaining only the terms with one quantum field gives (here we leave the trace over spin, 
space and time labels implicit, denote  spin matrix structure in bold and collapse the space and time indices):
\begin{widetext}
\begin{eqnarray}
S_{1,eq}^{\Sigma}&=&iTr\bigg[{\mathbf G}^R(1-2){\mathbf R}^\dagger_C(2)
\Sigma^K_{N-C}(2-1){\mathbf R}_Q(1)+{\mathbf G}^K(1-2){\mathbf R}^\dagger_C(2)
\Sigma^A_{N-C}(2-1){\mathbf R}_Q(1)
\nonumber \\
&+&{\mathbf G}^A(1-2){\mathbf R}^\dagger_Q(2)\Sigma^K_{N-C}(2-1){\mathbf R}_C(1)+
{\mathbf G}^K(1-2){\mathbf R}^\dagger_Q(2)\Sigma^R_{N-C}(2-1){\mathbf R}_C(1)\bigg]
\label{S1eqsigma}
\end{eqnarray}
\end{widetext}
Our assumption of spin-independent reservoirs means that $\Sigma^{R,A,K}$ are scalars in spin space so may be factored out, leaving

\begin{equation}
S_{1,eq}^\Sigma=i{\mathbf D}_{CQ}(1-2){\mathbf R}^\dagger_C(2){\mathbf R}_Q(1)+(Q\leftrightarrow C)
\label{S1eq1}
\end{equation}
Here,
\begin{equation}
{\mathbf D}_{CQ}={\mathbf G}^R(1-2)\Sigma^K_{N-C}(2-1)+{\mathbf G}^K(1-2)\Sigma^A_{N-C}(2-1)
\end{equation}
and
\begin{equation}
{\mathbf D}_{QC}={\mathbf G}^A(1-2) \Sigma^K_{N-C}(2-1)+{\mathbf G}^K(1-2)\Sigma^R_{N-C}(2-1)
\end{equation}

Since $G^{R/A}$ and $\Sigma^{R/A}_{N-C}$ are short ranged in space and time we can make a gradient expansion of the product of rotation matrices, writing ${\mathbf R}^\dagger_C(2)={\mathbf R}^\dagger_C(1)+(2-1)\partial {\mathbf R}^\dagger_C(1)+{\mathcal O}(1-2)^2$ and ${\mathbf R}_C(1)={\mathbf R}_C(2)+(1-2)\partial {\mathbf R}^\dagger_C(2)+{\mathcal O}(1-2)^2$. We then change labels $(1\leftrightarrow 2)$ in the second term and obtain
\begin{eqnarray}
&&\int d2 \left[{\mathbf D}_{CQ}(1-2){\mathbf R}^\dagger_C(1){\mathbf R}_Q(1)\right. \nonumber \\
&&\left. +{\mathbf D}_{QC}(1-2)
{\mathbf R}^\dagger_Q(1){\mathbf R}_C(1)\right]
\end{eqnarray}
Since $\int d2 D_{CQ}(1-2)=\int d2 D_{QC}(1-2)$, the leading term vanishes while the even parity of the equilibrium state  means  the first space derivative vanishes. So the leading term is the first time derivative and incorporating the $i$ into the coefficients we obtain
\begin{equation}
S_{1,eq}^\Sigma=Tr\left[{\mathbf D}_{CQ}^{'}\partial_t{\mathbf R}^\dagger_C(1){\mathbf R}_Q(1)
+{\mathbf D}_{QC}^{'}{\mathbf R}^\dagger_Q(1)\partial_t{\mathbf R}_c(1)\right]
\end{equation}
with
\begin{eqnarray}
{\mathbf D}_{CQ}^{'}&=&i\int d2 (t_2-t_1){\mathbf D}_{CQ}({\vec r}_1-{\vec r}_2,t_1-t_2)
\\
{\mathbf D}_{QC}^{'}&=&i\int d2 (t_1-t_2){\mathbf D}_{QC}({\vec r}_1-{\vec r}_2,t_1-t_2)
\end{eqnarray}
The coefficients are most conveniently evaluated in Fourier space as ${\mathbf D}_{CQ}^{'}=\partial {\mathbf D}_{CQ}(\Omega)/\partial \Omega|_{\Omega=0}$ and ${\mathbf D}_{QC}^{'}=-\partial {\mathbf D}_{QC}(\Omega)/\partial \Omega|_{\Omega=0}$. Introducing
\begin{eqnarray}
 {\mathbf G}^{'}&=& \frac{1}{2}\left({\mathbf G}^A(1-2)+{\mathbf G}^R(1-2)\right)
\\
{\mathbf A}&=& \frac{1}{2i}\left({\mathbf G}^A(1-2)-{\mathbf G}^R(1-2)\right)
 \\
 \Sigma^{'}_{N-C}&=& \frac{1}{2}\left(\Sigma^A_{N-C}(1-2)+\Sigma^R_{N-C}(1-2)\right)
 \\
 \Gamma_{N-C}&=& \frac{1}{2i}\left(\Sigma^A_{N-C}(1-2)-\Sigma^R_{N-C}(1-2)\right)
 \end{eqnarray}
and noting that in equilibrium we have  $G^K(k,\omega)=-2iA(k,\omega)h(\omega)$ and $\Sigma^K_{N-C}(k,\omega)=-2i\Gamma_{N-C}(k,\omega)h(\omega)$ with $h$ the equilibrium distribution function we find
 \begin{eqnarray}
 D_{CQ}(\Omega)&=&D_{1}(\Omega)-iD_{2}(\Omega)
 \\
 D_{QC}(\Omega)&=&-D_{1}(\Omega)-iD_{2}(\Omega)
 \end{eqnarray}
 with the real functions $D_1$ and $D_2$ given by
\begin{eqnarray}
 D_1(\Omega)&=&2\int(dkd\omega){\mathbf A}(k,\omega+\Omega)\Gamma_{N-C}(k,\omega)
  \nonumber\\
 &&\hspace{0.7in}\times\left(h(\omega+\Omega)-h(\omega)\right)
 \end{eqnarray}
 and
\begin{eqnarray}
D_2(\Omega)&=&2\int(dkd\omega) \left[{\mathbf G}^{'}(k,\omega+\Omega)\Gamma_{N-C}(k,\omega)h(\omega)
\right.
\nonumber \\
&&\left. \hspace{0.2in}+{\mathbf A}(k,\omega+\Omega)\Sigma_{N-C}^{'}(k,\omega)h(\omega+\Omega)
\right]
\end{eqnarray}
Thus, $D_{CQ}(\Omega=0)=D_{QC}(\Omega=0)$ as noted above. We therefore obtain
\begin{eqnarray}
&&S_{1,eq}^\Sigma=\sum_i\int dt
\nonumber \\
&&{\mathbf D_{berry}}\left(-i\partial_t{\mathbf R}^\dagger_C(i,t){\mathbf R}_Q(i,t)
+{\mathbf R}^\dagger_Q(i,t)i\partial_t{\mathbf R}_C(i,t)\right)
\nonumber
\\
&&+{\mathbf D_{diss}}\left(\partial_t{\mathbf R}^\dagger_C(i,t){\mathbf R}_Q(i,t)
+{\mathbf R}^\dagger_Q(i,t)\partial_t{\mathbf R}_C(i,t)\right)
\end{eqnarray}
with
\begin{equation}
{\mathbf D_{diss}}=\frac{\partial D_1}{\partial \Omega}=\frac{2}{\pi}\int(dk){\mathbf A}(k,\omega=0)
\Gamma_{N-C}(k,\omega=0)
\label{Ddiss}
\end{equation}
and, using the Kramers-Kronig relation and shifting $\omega \rightarrow \omega-\Omega$ appropriately,
\begin{eqnarray}
&&{\mathbf D_{berry}}=-2\int (d\omega dx)(dk) \nonumber \\
&&\frac{{\mathbf A}(k,x)\Gamma_{N-C}(k,\omega)\left(h(\omega)-h(x)\right)}{(\omega-x)^2}
\end{eqnarray}
The ${\mathbf D_{berry}}$ term appears as a renormalization of Eq.~(\ref{SG0}) while the ${\mathbf D_{diss}}$ term parametrizes effects of  dissipation. In the rotated basis both terms are diagonal matrices in spin space and are explicitly proportional to the spin relaxation rate $\Gamma_{N-C}$. Additional terms would appear if the reservoir was magnetic.

We have written Eq.~(\ref{Ddiss}) for the general case. Our specific results are presented for a momentum-independent spin relaxation rate. For this case the $\Gamma_{N-C}$ may be removed from the integral and we identify $\int (dk) A(k,\omega=0)$ as the spin-dependent Fermi level
density of states $N_\sigma$. In a simple model with $k^2/2m$ dispersion, $N_\sigma=mk_{F,\sigma}/(2\pi^2)$.
Thus we estimate that the diagonal matrix elements $D^\sigma$ are
\begin{equation}
D_{diss}^\sigma=\frac{3}{\pi}\frac{n_\sigma\Gamma_{N-C}}{E_{F,\sigma}}
\label{Ddiss1}
\end{equation}
with $E_{F,\sigma}=k_{F,\sigma}^2/2m$ the Fermi energy for spin direction $\sigma$.

\subsection{$S_2^{eq}$}
Expanding $S_2^{eq}$, Eq.~(\ref{S2eqdef}), gives the standard spin-stiffness terms, renormalized  weakly by the coupling to the reservoir. These may be written (reverting here to the physical spin coordinates)
\begin{equation}
S_{stiffness}=i\rho_S\int dx dt \,\,{\hat {\mathbf n}}_q\cdot {\vec \nabla^2}{\hat {\mathbf n}}_{cl,\perp}(x,t)
\label{Sstiffness}
\end{equation}
where ${\hat {\mathbf n}}_{cl,\perp}(x,t)$ is the classical component of ${\hat n}$ perpendicular to  the local magnetization direction and $\rho_S$ is the spin stiffness, proportional to the local magnetization.

\subsection{Nonequilibrium terms}

We  first consider the first term in  Eq.~(\ref{Sneqdef}). Expanding as in the previous section we find  terms of the form of  Eq.~(\ref{S1eq0}) but with ${\mathbf G}^K$ replaced by the nonequilibrium Keldysh function and terms of the form of Eq.~(\ref{S1eqsigma}) with the $G^{R,A}$ terms removed  and the ${\mathbf G}^K$ replaced by the nonequilibrium Keldysh function. The  terms of the form of  Eq.~(\ref{S1eq0})  are
\begin{eqnarray}
S_{2}^{neq,G_0}&=&-iTr\bigg[{\mathbf G}^K_{neq}(1-2){\mathbf R}^\dagger_C(2)
G_0^{A,-1}(2-1){\mathbf R}_Q(1)
\nonumber \\
&+&{\mathbf G}^K _{neq}(1-2){\mathbf R}^\dagger_Q(2)G_0^{R,-1}(2-1){\mathbf R}_C(1)\bigg]
\label{S1neq0}
\end{eqnarray}
Since the nonequilibrium ${\mathbf G}^K_{neq}$ is an odd function of its space coordinates but an even function of time the only non-vanishing contribution comes from the  hopping terms in $\delta {\mathbf H}$. We obtain
\begin{eqnarray}
&&S_2^{neq,G_0}=iTr\bigg[{\mathbf G}_{neq}^K(i,j;t)t_{j-i}
\nonumber
\\
&&\times\left({\mathbf R}^\dagger_C(j,t){\mathbf R}_Q(i,t)+{\mathbf R}^\dagger_Q(j,t)
{\mathbf R}_C(i,t)\right)\bigg]
\label{S2neqdef2}
\end{eqnarray}
Writing ${\mathbf R}^\dagger_C(j)=({\vec r}_j-{\vec r}_i)^a\nabla ^a{\mathbf R}^\dagger_C(i)$ and ${\mathbf R}_C(i)=({\vec r}_i-{\vec r}_j)^a\nabla ^a{\mathbf R}_C(j)$, changing $i$ to $j$ in the second term and
using Eqs.~(\ref{spincurrent}),~(\ref{Jdef}) we obtain
\begin{eqnarray}
&&S_{2}^{neq,G_0}=
-iTr\bigg[{\mathbf J}^a\left(\nabla^a{\mathbf R}^\dagger_C(i,t){\mathbf R}_Q(i,t)
\right. \nonumber \\
&&\left. -{\mathbf R}^\dagger_Q(i,t)\nabla^a{\mathbf R}_C(i,t)\right)\bigg]
\label{S2neqdef3}
\end{eqnarray}

We next consider the terms arising from ${\mathbf \Sigma}_{eq}$.  These are of the form of Eq.~(\ref{S1eqsigma}) but with the functions $D_{QC/CQ}$ replaced by
\begin{equation}
E_{CQ}={\mathbf G}^K_{neq}(1-2) \Sigma^A_{N-C}(2-1)
\end{equation}
and
\begin{equation}
E_{QC}={\mathbf G}^K_{neq}(1-2) \Sigma^R_{N-C}(2-1)
\end{equation}
Making the gradient expansion as above gives
\begin{equation}
Tr\left[{\mathbf E}_{CQ}^{a}\nabla^a{\mathbf R}^\dagger_C(1){\mathbf R}_Q(1)
+{\mathbf E}_{QC}^{a}{\mathbf R}^\dagger_Q(1)\nabla^a{\mathbf R}_c(1)\right]
\end{equation}
with
\begin{eqnarray}
{\mathbf E}_{CQ}^{a}&=&i\int d2 ({\vec r}_2-{\vec r}_1)^a{\mathbf E}_{CQ}({\vec r}_1-{\vec r}_2,t_1-t_2)
\\
{\mathbf E}_{QC}^{'}&=&i\int d2 ({\vec r}_1-{\vec r}_2)^a{\mathbf E}_{QC}({\vec r}_1-{\vec r}_2,t_1-t_2)
\end{eqnarray}
In the limit of momentum-independent (space local) spin relaxation rates which we consider here these terms vanish, but in general there will be a contribution proportional to $\partial_k\Sigma_{N-C}$.

In the  second term in Eq.~(\ref{Sneqdef}),  use of the steps leading to Eq.~(\ref{S1eqsigma})
along with the fact that $\Sigma_{neq}$ has only a Keldysh component and is proportional to the unit tensor in spin space gives (leaving the sum over coordinates implicit)
\begin{eqnarray}
S_{1,neq}^\Sigma&=&i{\mathbf F}_{CQ}(1-2){\mathbf R}^\dagger_C(2){\mathbf R}_Q(1)
\nonumber\\
&+&i{\mathbf F}_{QC}(1-2){\mathbf R}^\dagger_Q(2){\mathbf R}_C(1)
\label{S1neq1}
\end{eqnarray}
with
\begin{equation}
{\mathbf F}_{CQ}={\mathbf G}^R(1-2)\Sigma^K_{neq}(2-1)
\end{equation}
and
\begin{equation}
{\mathbf F}_{QC}={\mathbf G}^A(1-2) \Sigma^K_{neq}(2-1)
\end{equation}
Again making the gradient expansion and incorporating the $i$ into the coefficients we find
\begin{equation}
\delta S_{1,neq,\Sigma}^\Sigma= {\mathbf F}^{a}_{CQ} \nabla^a{\mathbf R}^\dagger_C(1){\mathbf R}_Q(1)
+ {\mathbf F}_{QC}^{a}{\mathbf R}^\dagger_Q(1) \nabla^a{\mathbf R}_C(1)
\label{SneqQC2}
\end{equation}

The explicit form of $\Sigma^K_{neq}$  (Eq.~(\ref{SigmaneqKfinal}))  gives, in a mixed position-frequency representation
\begin{eqnarray}
{\mathbf F}^{a}_{CQ}&=&\frac{i}{\pi}\sum_{j} {\mathbf G}^R(i-j,\omega=0))
t_{j-i}{\vec r}_{ji}^a\left({\vec r}_{ji}\cdot{\vec E}\right)\label{FCQa}
\\
{\mathbf F}^{a}_{QC}&=&\frac{i}{\pi}\sum_{j} {\mathbf G}^A(i-j,\omega=0))t_{j-i}{\vec r}_{ij}^a
\left({\vec r}_{ji}\cdot{\vec E}\right)\label{FQCa}
\end{eqnarray}
Here ${\vec r}_{ij}$ is the vector connecting site $i$ to site $j$ and in the $QC$ term we have ${\vec r}_{ij}^b$ not ${\vec r}_{ji}^b$ because we put the gradient on the $R_C$. In Fourier space we have
\begin{eqnarray}
{\mathbf F}^{a}_{CQ}&=&
\frac{i}{\pi}\int(dk) {\mathbf G}^R(k,\omega=0)\frac{\partial^2\varepsilon_k}
{\partial k_a\partial k_b}E^b\label{FCQa1}
\\
{\mathbf F}^{a}_{QC}&=&-\frac{i}{\pi}\int(dk) {\mathbf G}^A(k,\omega=0)
\frac{\partial^2\varepsilon_k}{\partial k_a\partial k_b}E^b\label{FQCa1}
\end{eqnarray}

Our final expression therefore is
\begin{eqnarray}
&&S_{1,neq}^{\Sigma}=\sum_i\int dt
\nonumber \\
&&i{\mathbf F}_{J}^a\left(-\nabla^a{\mathbf R}^\dagger_C(i,t){\mathbf R}_Q(i,t)+
{\mathbf R}^\dagger_Q(i,t)\nabla^a{\mathbf R}_C(i,t)\right)
\nonumber
\\
&&+{\mathbf F}^a_{diss}\left(\nabla^a{\mathbf R}^\dagger_C(i,t){\mathbf R}_Q(i,t)
+{\mathbf R}^\dagger_Q(i,t)\nabla{\mathbf R}_C(i,t)\right)\nonumber \\
\end{eqnarray}
with the ${\mathbf F}$  diagonal matrices in spin space with up and down spin components (here we use 
the resistivity to express the electric field in terms of the current)
\begin{eqnarray}
&&F^{a}_{J,\sigma}=-\frac{1}{\pi}\int(dk){\mathcal Re} G_\sigma^R(k,\omega=0)
\frac{\partial^2\varepsilon_k}{\partial k_a\partial k_b}\rho_\sigma J^b_\sigma\label{FJa}
\\
&&F^a_{diss,\sigma}=\frac{1}{\pi}\int(dk) A_\sigma(k,\omega=0)\frac{\partial^2
\varepsilon_k}{\partial k_a\partial k_b}\rho_\sigma J^b_\sigma\label{Fdiss}
\end{eqnarray}

It will be useful for  our further discussion to estimate the magnitudes of $F^a$ in a simple model with $k^2/2m$ dispersion. In this case $\partial^2\varepsilon_k/\partial k_a\partial k_b=\delta^{ab}/m$  and the estimates that led to Eq.~(\ref{Ddiss1}) along with Eq.~(\ref{rhosimple}) give
\begin{equation}
F^a_{diss,\sigma}=\frac{3\Gamma^{tot}_\sigma}{2\pi E_{F,\sigma}} J^a
\label{Fdiss1}
\end{equation}

\section{Second order in quantum fields: Noise terms \label{Noise}}

Expanding the action to second order in quantum fields yields terms which stabilize the integral over quantum fields and, among other effects, introduce an effective temperature  into the problem.  A general analysis of these terms is somewhat involved; we focus here  on obtaining  the qualitative behavior at $T$=$0$ and to leading  order in the gradient expansion.  In this case the equilibrium terms do not contribute and expanding  Eq.~(\ref{Sneqdef}) to second order in ${\mathbf R}_Q$ yields
\begin{equation}
S_{noise}=iTr\left[{\mathbf G}^K_{eq}{\mathbf R}^\dagger_Q{\mathbf \Sigma}^K_{neq}{\mathbf R}_Q+{\mathbf
G}^K_{neq}{\mathbf R}^\dagger_Q{\mathbf\Sigma^K_{eq}}{\mathbf R}_Q\right]
\label{Snoise}
\end{equation}
For simplicity we assume that the spin-nonconserving scattering rate is small compared to the total scattering rate, so that the first term is dominant. We then obtain
\begin{equation}
S_{noise}=iT^*\sum_i\int dtTr\left[{\mathbf N}{\mathbf R}^\dagger_Q(i,t){\mathbf \Gamma^{tot}}{\mathbf R}_Q(i,t)\right]
\label{Snoise1}
\end{equation}
${\mathbf N}$ is a diagonal matrix with entries $N_\sigma=\frac{2}{\pi}\int (dk)|cos \phi_k|A_\sigma(k,\omega$=$0)$, $\phi_k$ is the angle between ${\vec v}_k$ and ${\vec E}$, ${\mathbf \Gamma}$ is a diagonal matrix with entries $\sqrt{\Gamma^{tot}_\sigma\Gamma^{tot}}$
with
\begin{equation}
\Gamma^{tot}=\frac{\Gamma^{tot}_\uparrow+\Gamma^{tot}_\downarrow}{2}
\end{equation}
and the effective temperature is
\begin{equation}
T^*=\frac{|{\vec E}||{\vec v}_F|}{\sqrt{\Gamma_{N-C}\Gamma^{tot}}}
\label{Tstardef}
\end{equation}
As in Ref.~\onlinecite{Mitra08}, the effective temperature is given by the voltage drop over the  distance traveled between inelastic scattering events.

\section{Summary of Action \label{Action}}

Collecting all of the terms we obtain an action in terms of the classical and quantum rotation fields of the form
\begin{equation}
S_F\left[\{{\mathbf R}^\dagger,{\mathbf R}\}\right]=S_B+S_{diss}+S_{stiffness}+S_{noise}
\label{Sfinal}
\end{equation}
with
\begin{equation}
S_B=\sum_\mu Tr\left[{\cal {\mathbf B}}_\mu\Bigl\{{\mathbf R}^\dagger_Q\left(i\partial_\mu{\mathbf R}_C\right)
-\left(i\partial_\mu{\mathbf R}^\dagger_C\right){\mathbf R}_Q\Bigr\}\right],
\label{SB}
\end{equation}
\begin{equation}
S_{diss}=\sum_\mu Tr\left[{\cal {\mathbf D}}_\mu\Bigl\{{\mathbf R}^\dagger_Q\left(\partial_\mu{\mathbf R}
_C\right)+\left(\partial_\mu{\mathbf R}^\dagger_C\right){\mathbf R}_Q\Bigr\}\right]
\label{Sdiss}
\end{equation}
$S_{stiffness}$ is given by Eq.~(\ref{Sstiffness}) and $S_{noise}$ by Eq.~(\ref{Snoise1}). Here $\mu$=$0$ denotes the time direction and $\mu$=$a$ a space direction.

We have
\begin{eqnarray}
{\mathbf B}_0&=&m\sigma_z+{\mathbf D}_{berry}
\\
{\mathbf B}_a&=&{\mathbf J}^a+{\mathbf F}_J^a
\\
{\mathbf D}_0&=&{\mathbf D_{diss}}
\\
{\mathbf D}_a&=&{\mathbf F}_{diss}^a
\end{eqnarray}
Bold-face coefficients denote matrices in spin space; the spin structure arises from spin  dependence of the density of states and scattering rate in the presence of a non-vanishing local magnetization.

We now use Eq.~(\ref{omegadef}) to  express Eq.~(\ref{Sdiss})  in terms of the classical and quantum components of the unit vectors ${\hat \Omega}$. We have
\begin{equation}
{\mathbf R}_Q={  \hat\Omega}_Q^z{\mathbf 1} +i{\hat z}\times {  \hat\Omega}_Q^\bot\cdot {\vec \sigma}
\end{equation}
and
\begin{equation}
\partial_\mu{\mathbf R}_C=\frac{\partial {\mathbf R}_C}{\partial \Omega^b}\partial_\mu {\hat \Omega^b}
\end{equation}
The action is thus of the form
\begin{eqnarray}
&&S_F=\Omega^a_QJ^{ab}_\mu \partial_\mu \Omega^b_C+\Omega_Q^aD^{ab}_\mu \partial_\mu \Omega^b_C\nonumber \\
&&+ i T^* \Omega_Q^aN^{ab}\Omega_Q^b+S_{stiffness}
\label{Gterms1}
\end{eqnarray}
with
\begin{equation}
J^{ab}_\mu=iTr\left[{\mathcal {\mathbf B}}_\mu\left(\frac{\partial {\mathbf R}^\dagger_C}{\partial {\hat \Omega}^a}\frac{\partial {\mathbf R}_C}{\partial {\hat \Omega}^b}-\frac{\partial {\mathbf R}^\dagger_C}{\partial {\hat \Omega}^b}\frac{\partial {\mathbf R}_C}{\partial {\hat \Omega}^a}\right)\right]
\label{Jabdef}
\end{equation}
\begin{equation}
D^{ab}_\mu=Tr\left[{\mathcal {\mathbf D}}_\mu\left(\frac{\partial {\mathbf R}^\dagger_C}{\partial {\hat \Omega}^a}\frac{\partial {\mathbf R}_C}{\partial {\hat \Omega}^b}+\frac{\partial {\mathbf R}^\dagger_C}{\partial {\hat \Omega}^b}\frac{\partial {\mathbf R}_C}{\partial {\hat \Omega}^a}\right)\right]
\label{Dabdef}
\end{equation}
and
\begin{eqnarray}
N^{zz}&=&2\frac{N_\uparrow\Gamma^{tot}_\uparrow+N_\downarrow\Gamma^{tot}_\downarrow}{2}
\\
N^{xx}=N^{yy}&=&2\frac{N_\uparrow\Gamma^{tot}_\downarrow+N_\downarrow\Gamma^{tot}_\uparrow}{2}
\end{eqnarray}

Evaluating the derivatives explicitly and noting that   $J^{ab}$ and $D^{ab}$ have terms $\sim {\mathbf 1}$ and $\sim{\mathbf \sigma}_z$ we find
\begin{equation}
J_\mu^{xy}=-J_\mu^{yx}=2{\cal J}_\mu
\end{equation}
with all other elements of $J_\mu$=$0$, while
\begin{equation}
D^{ab}_\mu={\cal D}_\mu\delta^{ab}
\end{equation}

We thus have
\begin{equation}
S_B+S_{diss}  ={\hat \Omega}_Q\cdot \left(-{\hat z} \times \left({\cal J}_\mu\partial_\mu\right){\hat  \Omega}_C+\left({\cal D}_\mu \partial_\mu\right){\hat  \Omega}_C\right)
\label{Sfinal1}
\end{equation}
with $\mu=0,a$ labeling space-time components and
\begin{eqnarray}
{\cal J}_0&=&m+2Tr\left[\sigma_z{\mathbf D}_{berry}\right]
\\
{\cal J}_a&=&2Tr\left[\sigma_z\left({\mathbf J}^a+{\mathbf F}_J^a\right)\right]
\\
{\cal D}_0&=&2Tr\left[{\mathbf D}_{diss}\right]
\\
{\cal D}_a&=&2Tr\left[{\mathbf F}^a_{diss}\right]
\end{eqnarray}
The estimates provided above imply
\begin{eqnarray}
{\cal J}_0&\sim&m
\\
{\vec {\cal J}}&\sim&{\vec J}_M\sim m{\vec J}
\\
{\cal D}_0&\sim&\frac{\Gamma_{N-C}}{E_F}
\\
{\vec {\cal D}}&\sim&\frac{\Gamma^{tot}}{E_F}{\vec J}
\end{eqnarray}

The coefficient of $\Omega_Q$ in Eq.~(\ref{Sfinal1}) gives the classical equation of motion for an  isolated spin in the presence of dissipation and current drive. It is consistent  with previous derivations based  on the phenomenological Landau-Lifshitz-Gilbert equation,~\cite{Thiele73,Zhang04,Thiaville05} and with  microscopic derivations of the Landau-Lifshitz equations in the presence of current.~\cite{Duine07,MacDonald09} Here the terms proportional to ${\cal J}$  are non-dissipative terms giving rise to the adiabatic spin torque and arising from the usual Berry phase,  as modified by an applied current,~\cite{Bazaliy98,Shibata05,Shibata06} while the terms involving ${\cal D}$ arise from dissipative processes and give rise
to the nonadiabatic spin torque.~\cite{Zhang04}

The Galilean transformation
\begin{equation}
{\vec r}\rightarrow {\vec r}-{\vec u}_st
\end{equation}
with
\begin{equation}
{\vec u}_s=\frac{{\vec {\cal J}}}{{\cal J}_0}
\label{usdef}
\end{equation}
shifts to a frame co-moving with the spin current. In this reference frame the  spatial derivative  terms in $S_B$ 
vanish. The coefficient of the spatial derivative in the combination of $S_{diss}$ becomes
\begin{equation}
{\cal D}_0\left({\vec u}_n-\vec{u}_s\right)
\label{Galilean}
\end{equation}
with normal velocity
\begin{equation}
{\vec u}_n=\frac{{\vec {\cal D}}}{{\cal D}_0}
\label{undef}
\end{equation}
In a Galilean invariant theory this combination vanishes. The easiest way to achieve Galilean invariance is to remove the scattering, so ${\cal D}_\mu$=$0$, however one may also imagine a situation in which ${\vec u}_n={\vec u}_s$. However, the estimates given above show  that in the general case, in the co-moving frame the coefficient of the spatial derivative in the  ${\cal D}$ term is of order of the difference between the transport and spin relaxation rates  (measured relative to the Fermi energy)
\begin{equation}
\frac{\Gamma_{N-C} -\Gamma^{tot}}{E_F}{\vec J}
\end{equation}
Because we generically expect a spin relaxation rate much less than the transport relaxation rate, Galilean invariance is strongly broken in general.

\section{Equation of Motion for a  Topological Defect\label{Topological}}

\subsection{Overview}

To obtain the equation of motion of a vortex we imagine that spin wave excitations can be integrated out so that the spin orientation ${\mathbf n}$ is determined by topological defects at positions ${\vec X}^j(t)$.  We further suppose that the topological defects move slowly enough that the spins are  relaxed to the values appropriate to the instantaneous position of the defects.  We have (repeated indices summed)
\begin{equation}
{\hat \Omega}_Q=\frac{\delta {\hat \Omega}}{\delta X^a_{C,j}}X^a_{Q,j}
\end{equation}
Then we may rewrite Eq.~(\ref{Sfinal1}) as
\begin{eqnarray}
S&=&\int dt\sum_{ia}  X^a_{Q,i}(t){\cal F}^a_i(t)+\sum_{ijab}X^a_{Q,i}(t){\cal L}_{ij}^{ab}(t){\dot X}^b_{C,j}(t)
\nonumber \\
&&\hspace{0.9in}+i\sum_{ij}X^a_{Q,i}(t)\Xi^{ab}X^b_{Q,j}(t)
\label{SX1}
\end{eqnarray}
Here the noise term
\begin{equation}
\Xi^{ab}=T^*\frac{\delta {\hat \Omega^{\mu}}}{\delta {X^a_{C,i}}}N^{\mu\nu}
\frac{\delta {\hat \Omega^{\nu}}}{\delta { X^b_{C,j}}}
\label{Xiab}
\end{equation}
The generalized force ${\cal F}$ and Liouville operator ${\cal L}$  are formally given by

\begin{eqnarray}
{\cal F}^a_i(t)  &=&\int (dr){\hat z}\cdot\frac{\delta {\hat \Omega}}{\delta
X^a_{C,i}}\times\left({\vec {\cal J}}\cdot {\vec \nabla}{\hat  \Omega}\right)
\nonumber\\
&&+\frac{\delta {\hat \Omega}}{\delta X^a_{C,i}}\cdot\left({\vec {\cal D}}\cdot {\vec \nabla}{\hat  \Omega}\right)-\sum_j\frac{\partial U({\vec X}_{C,i}-{\vec X}_{C,j})}{\partial  X^a_{C,i}}\nonumber \\\label{Fdef}
\end{eqnarray}
\begin{equation}
{\cal L}_{ij}^{ab}(t)=\int (dr){\cal J}_0{\hat z}\cdot\frac{\delta {\hat \Omega}}
{\delta X^a_{C,i}}\times\frac{\delta {\hat \Omega}}{\delta X^b_{C,j}}+{\cal D}_0
\frac{\delta {\hat \Omega}}{\delta  X^a_{C,i}}\cdot\frac{\delta {\hat \Omega}}{\delta  X^b_{C,j}}
\label{Ldef}
\end{equation}

Here ${\hat \Omega}$, ${\hat z}$ denote  unit vectors in spin space. They depend on all of the vortex positions, although this dependence is not explicitly notated. The ${\vec \nabla}$ are  vectors in position space. Except for the terms in parentheses in Eq.~(\ref{Fdef}) the dot and cross product structure refers to the spin directions. ${\vec {\cal J}}$ is the spin current produced by an applied electric field. The direction of current flow (indicated by the vector symbol) is along the applied  electric field while the spin direction is parallel to the local magnetization. ${\vec {\cal D}}$ is also directed along the applied electric field.  $U$ is the interdefect potential arising from the spin stiffness and dipolar forces.

The equation of motion for vortex $i$ is obtained in the usual way~\cite{Kamenevrev} by using a Lagrange multiplier to  integrate out the quantum fields, leading to
\begin{equation}
\eta {\cal C}{\cal J}_0 {\hat z} \times {\dot {\vec X}}+I_D {\cal D}_0\dot {\vec X}_i=\eta {\cal C} {\hat z} \times {\vec {\cal J}}+I_D{\vec {\cal D}}+\sum_j{\vec F}_{ij}+{\vec \zeta}_i
\label{vortexEOM}
\end{equation}
with ${\vec F}_{ij}=-{\vec \nabla}_{ij}U({\vec X}_{i}-{\vec X}_{j})$, $\eta=\pm1$ the sign of the vorticity and  $\zeta_i$ a fluctuating field which on scales longer than the mean free path may be taken to be delta correlated, with
\begin{equation}
\left< \zeta^a_i (t)\zeta^b_j(t{'}\right>=2\Xi^{ab}_{ij}\delta(t-t{'})
\label{noisecorr}
\end{equation}
Eq.~(\ref{vortexEOM}) is consistent  with derivations based on the phenomenological
Landau-Lifshitz-Gilbert equation.~\cite{Thiele73,Thiaville05}

In the remainder of this section we examine separately the drive terms (those arising due to the presence of an imposed current and visible in the equation of motion for a single topological defect) and the intervortex interaction terms.

\subsection{Drive terms}

We examine Eq.~(\ref{Ldef})  for a two dimensional system containing a single vortex, at a position ${\vec X}(t)=X(t){\vec e}_x+Y(t){\vec e}_y$. In this case $\Omega\rightarrow \Omega({\vec r}-{\vec X}(t))$ so $\delta {\hat \Omega}/\delta X^a=-\nabla^a{\hat \Omega}$ and Eqs.~(\ref{Fdef}),~(\ref{Ldef}) involve two quantities:
\begin{eqnarray}
I_J&=&\int d^2r {\cal M}(r)\left(\nabla_x {\hat \Omega}_x\nabla_y
{\hat \Omega}_y- \nabla_x {\hat \Omega}_y\nabla_y {\hat \Omega}_x\right)
\label{IJdef}
\\
&&I_D^{ab}=\int d^2r \nabla_a {\hat \Omega}\cdot \nabla_b {\hat \Omega}
\end{eqnarray}
with ${\cal M}$  proportional to the magnetization a function to be discussed below. $I_J$ and $I_D$  depend on the structure of the vortex. We consider here an easy-plane system with spins lying preferentially in the $x-y$ plane and choose the origin of coordinates to be the position of the vortex. Far from the vortex, ${\cal M}$ is a constant proportional to the local magnetization and the spin orientation vector is ${\hat n}_x=cos \eta \phi$, ${\hat n}_y=sin\eta\phi$ and ${\hat n}_z=0$, implying $\Omega_z=\frac{1}{\sqrt{2}}$ and $(\Omega_{x},\Omega_{y})=\frac{1}{\sqrt{2}}(cos\eta\phi,sin\eta\phi)$. Within a distance $\xi$ of the vortex the structure changes. The integral defining $I_D$ is dominated by the far region. Substituting we have
\begin{equation}
I_D^{ab}=\pi\eta^2\delta^{ab}ln\frac{L}{\xi}+...=I_D\delta^{ab}
\end{equation}
with the ellipsis denoting terms of order $1$. In a system with a single vortex the long distance  cutoff is the system size; in a system with equal densities of vortices and antivortices the cutoff  is the mean intervortex spacing. This logarithm has been previously noted.~\cite{Shibata06}

The noise term may be evaluated similarly. The leading behavior is also logarithmic and arises from the  case where $i$ and $j$ refer to the same vortex; we estimate $\Xi^{ab}_{ij}=\Xi^a\delta^{ab}_{ij}$ with
\begin{equation}
\Xi^{a}=I_DN^{aa}T^*
\label{noisecoeff}
\end{equation}
and $\Xi^z$ of order unity. Here $T^*$ is given in Eq.~(\ref{Tstardef}). Note that in a fully polarized (half-metallic) magnet, $N^{xx}=0$ and the dissipation vanishes in this approximation (higher order processes will lead to a non-zero, but much smaller, noise).

Turning now to $I_J$ we see that the contributions from the far region vanish so $I_J$ is determined by the properties of the vortex core, which depends upon details. Two simplifying limits are possible. In the extreme soft spin, high anisotropy limit, the core is defined by a vanishing of the local magnetization amplitude (represented here by ${\cal M}\rightarrow0$) while the spin vector remains in the easy plane. In this case $I_J$=$0$.  On the other hand, in the hard spin, weak anisotropy  limit the spin amplitude remains constant, and the spin  direction rotates out of the plane, becoming parallel (or anti-parallel) to $z$.   In a general situation both effects occur, with the relative importance controlled by the ratio of the magnetic anisotropy energy to the amplitude energy.

Going to polar ($r-\phi$) coordinates we have ${\hat \Omega}_x=sin\frac{\theta(r)}{2}cos\eta\phi$ and  ${\hat \Omega}_y=sin\frac{\theta(r)}{2}sin\eta\phi$ so
\begin{widetext}
\begin{eqnarray}
I_J=&&\int d^2r {\cal M}(r)\bigg\{\left(\frac{d\theta}{2dr}cos\frac{\theta}{2}cos^2\eta\phi +\frac{\eta}{r}sin\frac{\theta}{2}sin^2\eta\phi)\right)\left(\frac{d\theta}{2dr}cos\frac{\theta}{2}sin^2\eta\phi +\frac{\eta}{r}sin\frac{\theta}{2}cos^2\eta\phi)\right)
\nonumber \\
&&-\left(\frac{d\theta}{2dr}cos\frac{\theta}{2}cos\eta\phi sin\eta\phi -\frac{\eta}{r}sin\frac{\theta}{2}cos\phi sin\eta\phi)\right)\left(\frac{d\theta}{2dr}cos\frac{\theta}{2}cos\eta\phi sin\eta\phi -\frac{\eta}{r}sin\frac{\theta}{2}cos\eta\phi sin\eta\phi)\right)
\bigg\}
\end{eqnarray}
\end{widetext}
The terms which survive are
\begin{equation}
I_J=\eta\int d^2r {\cal M}(r)\frac{d\theta}{4rdr}sin\theta=\frac{\pi\eta}{2}\int dr {\cal M}(r)\frac{dcos\theta}{dr}
\end{equation}
At $r=\infty$ $cos\theta=0$ and at $r=0$ $cos\theta=\pm1$ according as the spin rotates parallel or antiparallel to the $z$ axis. Thus if ${\cal M}$ is constant $I_J$ is a topological invariant, proportional to the vorticity and to the sign of the rotation, but variation of the magnetization amplitude near the core means the amplitude is non-universal. In the extreme hard axis limit the spin remains in-plane over the whole region where ${\cal M}\neq 0$ and $I_J$ vanishes. In general we can write
\begin{equation}
I_J=\eta {\cal C}
\end{equation}
with ${\cal C}$ of order $1$ carrying the sign of the spin orientation in the vortex core and the magnitude factors associated with the  variation of ${\cal M}$ near the core.

Thus we finally obtain for the action
\begin{eqnarray}
S&=&i\int dt \sum_iT^*N^{xx}I_D{\vec X}_{Qi}\cdot {\vec X}_{Qi}
\nonumber\\
&&+\sum_iI_D{\vec X}_{Qi}\cdot\left({\cal {\vec D}}-{\cal D}_0{\dot{\vec X}}_{Ci}\right)
\nonumber \\
&&+ {\vec X}_{Qi}\cdot\left(\eta {\cal C}{\hat z}\times\left( {\vec {\cal J}}-{\cal J}_0{\dot{\vec X}}_{Ci}\right)\right)
\nonumber \\
&&-\sum_i{\vec X}_{Qi}\cdot\sum_j\frac{\partial U({\vec X}_{C,i}-{\vec X}_{C,j})}{\partial  {\vec X}_{C,i}}
\label{S1vortex}
\end{eqnarray}

With the help of Eq.~(\ref{usdef}) we identify
\begin{equation}
{\vec F}_M=\eta {\cal C}{\cal J}_0{\hat z}\times\left( \vec {u}_s-{\dot{\vec X}}\right)
\label{Fmagnus}
\end{equation}
as the Magnus force acting on a vortex. We see that it has a sign determined by the combination of the vorticity and the constant ${\cal C}$ which describes the out of plane direction of spin rotation in the vortex core. Comparison to the usual theory of neutral superfluids (see e.g. Eq.~(2.1) of Ref.~\onlinecite{Halperin80}) allows us to identify ${\cal J}_0$ with the combination $2\pi  \rho_S/m$ (essentially $2\pi$ times the density of superfluid particles). On the other hand, to make contact with results obtained for superconductors it is more useful to identify ${\cal {\vec J}}={\cal J}_0{\vec u}_s$ with the supercurrent and to note that the combination ${\cal J}_0{\dot{\vec X}}_C$ gives rise to the vortex Hall angle, which in superconductors is typically of order $T_c/E_F$ and is in most cases negligible.

Proceeding similarly and using Eq.~(\ref{undef}) we see that
\begin{equation}
{\vec F}_D=I_D{\cal D}_0\left({\vec u}_n-{\dot {\vec X}}\right)
\label{FD}
\end{equation}
represents the physics of dissipation.  A very similar term occurs  in superconductors and in the theory of neutral superfluid films  (see e.g. Eq.~(2.2) of Ref.~\onlinecite{Halperin80}), although in this case physics not relevant here produces an additional term proportional to ${\hat z}\times({\vec u}_n-{\dot {\vec X}})$.

In neutral superfluids and superconductors, ${\vec u}_n-{\dot {\vec X}}$ represents the  velocity of the vortex relative to the source of dissipation (the substrate, in the neutral superfluid case, or the normal-fluid excitations in a two-fluid model of the superconducting case). The relative magnitudes of $u_s$ and $u_n$ is an important issue in the theory of superfluids shaken by motion of their container, but except perhaps in the ultra-clean limit $u_n$ may be neglected in superconductors. However, in the magnetic case our estimates indicate that $u_s/u_n \sim \Gamma_{N-C}/\Gamma^{tot}$ is typically much less than unity.

The relative magnitudes of the dissipative ($I_D{\cal D}_0$) and non-dissipative (${\cal J}_0$) terms sets the vortex Hall angle. In typical superconducting films the dissipative terms are dominant, but in the magnetic situation of relevance here  the order of magnitude of the dissipative coefficient is set by the spin relaxation rate $\Gamma_{N-C}$ which is likely to be small relative to the other terms in the vortex equation of motion.

For a single vortex (so ${\vec F}$=$0$) and in the absence of noise we may write the solution of
Eq.~(\ref{vortexEOM}) explicitly as
\begin{eqnarray}
&&\frac{d \vec{X}}{dt} = \left(\frac{\eta {\cal C}{\cal J}_0I_D{\cal D}_0}
{\left(I_D{\cal D}_0\right)^2+\left(\eta {\cal C}{\cal J}_0\right)^2}\right)
\hat{z}\times \left(\vec{u}_s- \vec{u}_n\right) \nonumber \\
&&-
\left(\frac{\left(I_D{\cal D}_0\right)^2}{\left(I_D{\cal D}_0\right)^2+\left(\eta {\cal C}{\cal J}_0\right)^2}
\right)\left(\vec{u}_s - \vec{u}_n\right) + \vec{u}_{s}\label{magvorteom1}
\end{eqnarray}
We observe that if either the spin relaxation rate (and hence ${\cal D}_0$) vanishes or if   $\vec{u}_s =\vec{u}_n$,  Eq.~(\ref{magvorteom1})  implies, $\dot{X}= \vec{u}_s$, so that the vortex simply drifts with the applied current. Thus in order to get motion transverse to current flow one needs both non-zero spin relaxation and  a breaking of Galilean invariance $\vec{u}_{s} \neq \vec{u}_n$. In our model both of these effects arise from the presence of the spin reservoir.

Eq.~(\ref{magvorteom1}) is very similar to the equation of motion for a  vortex in  a neutral superfluid given in 
Eq.~(2.3) in Ref~\onlinecite{Halperin80} which we rewrite below,
\begin{eqnarray}
&&\frac{d \vec{R}}{dt} = \nonumber \\
&&2\pi n \hbar \rho_s^0/m\left(\frac{B}{((2\pi\hbar/m)\rho_s^0-B^{\prime})^2 + B^2}\right)
\hat{z}\times \left(\vec{v}_{n} - \vec{v}_s\right)\nonumber \\
&&+ C\left(\vec{v}_{n} - \vec{v}_s\right) + \vec{v}_{s} \label{sceomvor}\\
\nonumber
\end{eqnarray}
Here $\vec{v}_s$ is the superfluid velocity, while $\vec{v}_n$  is the velocity of an underlying substrate which is also responsible for the dissipation  and $n=\pm 1$ denotes the vorticity. $B$ is a  drag coefficient arising from coupling of the superfluid to the substrate, and $B{'}$ is a drag coefficient arising from coupling of ripplon excitations to the curl of the superfluid velocity which does not appear in the case considered here. When $B{'}=0$, the coefficient C defined in Ref.~\onlinecite{Halperin80}  equals $B^2/((2\pi\hbar/m)\rho_s^0)^2 + B^2)$. Identifying $B$ with $I_D{\cal D}_0$ and $(2\pi\hbar/m)\rho_s^0$ with ${\cal C}{\cal J}_0$ we see that the expressions become identical.

The important term for our subsequent considerations is the first one on the rhs of Eq.~(\ref{magvorteom1}), which shows that vortices have a component of motion which is transverse to an applied current with sign determined by the combination $\eta {\cal C}$. The estimates presented in the previous section show that ${\cal D}_0$ is of the order of the ratio of the spin relaxation rate $\Gamma_{N-C}$ to the fermi energy $E_F$ while ${\cal D}$ is of the order of the ratio of the total scattering rate $\Gamma^{tot}$ to the fermi energy. ${\cal J}_0$ is proportional to the magnetization while the spin current  ${\vec {\cal J}}$ is proportional to $m$ times the total current ${\vec J}$. Thus the transverse component $X_T$ of vortex motion is roughly
\begin{equation}
{\dot X}_T \sim -\left(\eta {\cal C}{\hat z}\times {\vec J}\right)\frac{ mI_D\frac{\Gamma^{tot}}{E_F}}{\left(I_D\frac{\Gamma_{N-C}}{E_F}\right)^2+(\eta m {\cal C})^2}
\end{equation}

Except very near to a magnetic transition or in the case of very strong spin relaxation the first term in the denominator is negligible and we have
\begin{equation}
{\dot X}_T \sim -\left(\eta {\cal C}{\hat z}\times {\vec J}\right)\left(\frac{ I_D\Gamma^{tot}}{m\eta^2 {\cal C}^2}\right)
\label{XT}
\end{equation}

\subsection{Interdefect Interaction}

The force ${\mathcal F}_{ij}$ between two vortices at positions $X_i$ and $X_j$ is given by
\begin{equation}
{\mathcal F}_{ij}=-\frac{\partial U({\vec X}_{C,i}-{\vec X}_{C,j})}{\partial  X^a_{C,i}}
\label{Fijdef}
\end{equation}
It has contributions from the dipole force and from the usual spin stiffness terms. Writing $U = U_{SS}+ U_{dip}$,  the interaction arising from the spin stiffness is
\begin{equation}
U_{SS}=\frac{\pi \rho_S}{2} \ln \left|\frac{{\vec X}_{12}}{d}\right|
\label{USS}
\end{equation}
with $\rho_S$ the spin stiffness per unit area.

In magnetic systems, unlike in superconductors, the dipole interaction may also be important. For a general spin configuration the dipolar energy may be written as
\begin{eqnarray}
U_{dip} = (g\mu_B)^2\int d^3x d^3x^{\prime}\frac{{\vec \nabla}\cdot {\vec  m}(x)
{\vec \nabla}\cdot {\vec  m}(x^{\prime})}
{|\vec{x}-\vec{x}^{\prime}|}
\end{eqnarray}
with $g$ the electron g-factor.  Specializing to  a two dimensional film of thickness $d$ and
considering the case of a vortex and an antivortex separated by the
vector ${\vec X}_{12}$ we obtain~\cite{Maier04}
\begin{eqnarray}
U_{dip} =\eta_1\eta_2E_{dip}({\hat X}_{12}\cdot {\hat m})\frac{X_{12}}{d}
\label{Udip}
\end{eqnarray}
Here the dependence of $X$ follows by power counting  and  $E_{dip}=(g\mu_B)^2d^3C_{dipole}({\hat X}_{12}\cdot {\hat m})$  is the dipolar energy associated with magnetization in a volume set by the film thickness $d$, multiplied by a coefficient which depends on the relative orientation of the vortex/antivortex separation and the magnetization. We have evaluated the constant   $C_{dipole}$ numerically, finding $C_{dipole}\approx 110$ for a vortex/antivortex pair separated along the direction defined by the magnetization and $C_{dipole}\approx 35$ for a vortex/antivortex pair separated perpendicular  to the direction defined by the magnetization.

\section{Current-driven vortex unbinding transition\label{Transition}}

The previous section showed that vortices of opposite vorticity tend to be driven apart by an applied spin current but are pulled together by the intervortex force. The competition between these two effects leads to a  current driven vortex unbinding transition. To investigate this possibility in more detail we consider the equation of motion of a single vortex-antivortex pair, obtained from the solutions of Eq.~(\ref{vortexEOM}) for two vortices of opposite vorticity 
(more precisely opposite sign of $\eta{\mathcal C}$, and in what follows we take $|\eta|=1$). 
We find that  the equation for intervortex distance  ${\vec X}_{12}$  can be written
\begin{equation}
{\dot {\vec X}}_{12}=A\left({\vec {\cal F}}-{\vec {\cal E}} \right)+{\vec {\cal \zeta}}
\label{transverseequation}
\end{equation}
with
\begin{equation}
A=\frac{2I_D {\cal D}_0}
{(I_D{\cal D}_0)^2+({\cal C}{\cal J}_0)^2}
\label{Adef}
\end{equation}

The force term coupling the two vortices is given by
Eq.~(\ref{Fijdef}). The current-induced force acting to separate the two vortices is
\begin{equation}
{\vec {\cal E}}={\cal C}{\hat z}\times
\left({\vec {\cal J}}-\frac{{\cal J}_0}{{\cal D}_0}{\vec {\cal D}}\right)
\end{equation}
We observe again that the net force vanishes if the Galilean-invariance condition ${\cal J}_0{\cal D}={\cal D}_0{\cal J}$ is satisfied.

$\zeta$ is  a noise field with correlations now given by
\begin{equation}
\left< \zeta^a_i (t)\zeta^b_j(t{'}\right>=2A T_{eff}\delta_{ij}\delta^{ab}\delta(t-t{'})
\label{noisecorr2}
\end{equation}
with
\begin{equation}
T_{eff}=\frac{N^{xx}T^*}{{\cal D}_0}
\label{Teffdef2}
\end{equation}

Eqs.~(\ref{transverseequation}),~(\ref{Udip}) show  that if the applied current is large enough that
\begin{equation}
\delta {\cal E}=\frac{{\cal C}}{{\cal D}_0}(-{\cal J}_0{\cal D}+{\cal D}_0{\cal J})-\frac{E_{dip}}{d}>0
\label{criticalcurrent}
\end{equation}
then at sufficiently long distances it becomes energetically favorable to separate vortex/antivortex pairs.

To calculate the probability of formation of vortex/antivortex pairs we follow 
Ref.~\onlinecite{Halperin80} and 
recast Eq.~(\ref{transverseequation})  as a Fokker-Planck equation for the probability $P$ per unit area of finding a vortex-antivortex pair at separation $r$:
\begin{equation}
\partial_tP({\vec r},t)=-{\vec \nabla}\cdot{\vec J}_P
\label{FPeq}
\end{equation}
Here, the  probability current $J_P$ is given by
\begin{equation}
{\vec J}_P=-A T_{eff}e^{-\frac{{\bar U}}{T_{eff}}}{\vec \nabla}P({\vec r},t)e^{\frac{{\bar U}}{T_{eff}}}
\label{JP}
\end{equation}
and effective potential
\begin{equation}
{\bar U}= \left[E_{dip}\frac{|{\vec r}|}{d}+\frac{\pi \rho_S}{2}
ln\frac{|{\vec r}|}{d}\right] - {\cal E} \hat{x}\cdot\vec{r}
\label{barU}
\end{equation}

In the limit of relatively weak applied currents, Eq.~(\ref{FPeq}) may be analyzed by saddle-point 
method.~\cite{Kramers40,Halperin80} We take the electrical current to be applied in the $y$ direction, so the vortex/antivortex pair separates along $x$ and choose $x=y=0$ as the location of the saddle point in ${\bar U}$ with $x\rightarrow \infty$ corresponding to large vortex/antivortex separation. We assume that the probability current density is only large near the saddle point of ${\bar U}$ and seek a steady-state solution with $J_P^y=0$. The pair creation rate per unit area  $R$ is then given in terms of the $y$ integral of the $x$ component of the probability current 
as~\cite{Kramers40,Halperin80}
\begin{equation}
R=\int dy J_P^x(x=0,y)
\label{Rdef}
\end{equation}

To determine $R$ we observe that the condition that $J_P^y=0$ is fulfilled if $P$ has a $y$ dependence which compensates the $y$ dependence of ${\bar U}$, i.e.
\begin{equation}
P({\vec r,t})e^{\frac{{\bar U}}{T_{eff}}}=f(x,t)
\label{fdef}
\end{equation}

For $x\rightarrow \infty$ the density of vortex/antivortex pairs is very low, so $f\rightarrow 0$, 
whereas  for intervortex separations much less than the saddle point value we expect the probability 
distribution to be the pseudoequilibrium one corresponding to temperature $T_{eff}$ so $f(x<<0)\rightarrow 1$.
The steady state condition is  fulfilled if $J^x_P$ depends only on $y$, i.e.
\begin{equation}
J_P^x=\Phi(y)=-A T_{eff}e^{-\frac{{\bar U}}{T_{eff}}}\partial_xf(x)
\label{Phidef}
\end{equation}

Multiplying both sides of Eq.~(\ref{Phidef}) by $e^{-\frac{{\bar U}}{T_{eff}}}$,  integrating over $x$ using the boundary conditions on $f$ and rearranging  then gives
\begin{equation}
\Phi(y)=\frac{AT_{eff}}{\int^\infty_{x_{low}}dx^{'}e^{\frac{{\bar U}(x^{'},y)}{T_{eff}}}}
\end{equation}
with $x_{low}$ sufficient far from the saddle point that $f\approx 1$.

To implement these formulae we now make the saddle point approximation explicit. The  saddle point location $r_c$ is
\begin{equation}
{\vec r}_c=\left(\frac{\pi \rho_S}{2\delta{\cal E}},0\right)
\end{equation}

Near the saddle point we write
\begin{equation}
\frac{{\bar U}(\delta x,\delta y)}{T_{eff}}={\bar u}+\frac{\delta y^2}{2y_c^2}-\frac{\delta x^2}{2x_c^2}
\end{equation}
with
\begin{eqnarray}
{\bar u}&=&\frac{\pi \rho_S}{2T_{eff}}ln\frac{r_c}{d}
-\frac{\pi \rho_S}{2T_{eff}}
\\
y_c^{-2}&=&\frac{E_{dip}}{T_{eff}d r_c}+\frac{\pi \rho_S}{2T_{eff}r_c^2}
\\
x_c^{-2}&=&\frac{\pi \rho_S}{2T_{eff}r_c^2}
\end{eqnarray}

Thus Eq.~(\ref{Phidef}) becomes
\begin{equation}
J_P^x(y)=\frac{AT_{eff}}{x_c\sqrt{2\pi}}e^{-{\bar u}}e^{-\frac{y^2}{2y_c^2}}
\end{equation}
so, performing the $y$ integral and restoring units
\begin{equation}
R=\frac{2I_D N^{xx}T^*}{(I_D{\cal D}_0)^2+({\cal C}{\cal J}_0)^2}
\frac{1}{\sqrt{1+\frac{E_{dip}}{d{\delta \cal E}}}}\left(\frac{2e\delta {\mathcal E}d}
{\pi \rho_S}\right)^{{\cal D}_0\frac{\pi \rho_S}{2N^{xx}T^*}}
\label{Rfinal}
\end{equation}

Because the vortex annihilation rate scales as the square of the vortex density and must balance the creation rate, we obtain
\begin{equation}
n_V\sim\left(\frac{2\delta {\mathcal E}d}
{\pi \rho_S}\right)^{{\cal D}_0\frac{\pi \rho_S}{4N^{xx}T^*}}
\end{equation}

\section{Conclusions \label{Conclusion}}

In this paper we have shown that an applied current can drive a nonequilibrium topological (defect-unbinding) transition in a continuous-symmetry magnet. We derived the appropriate action from microscopics and presented specific results for a quasi two-dimensional $XY$ symmetry magnet, providing estimates for the relevant length and current scales and characterizing the scaling of defect density as the current exceeds threshold. A qualitative result is that, in the presence of the dipole interaction, the critical current depends strongly (by a factor of $\sim 3$) on the angle between the current and the magnetization. Our general formalism and qualitative conclusions however apply to any combination of dimensionality and order parameter structure which supports topological defects. These results complement the treatment of current-driven quantum criticality studied in Refs.~\onlinecite{Mitra06,Mitra08}  which considered transitions dominated by current-induced local spin excitations in Ising and Heisenberg  magnets.

The basic physics driving the transition, namely that an applied current produces a force which acts oppositely on vortices of different chirality, is familiar from the theory of vortices in superconductivity~\cite{Halperin80,Minnhagen87} and has been derived~\cite{Thiaville05,Shibata05,Shibata06} from the Landau-Lifshitz-Gilbert equations which provide a phenomenological description of magnetization dynamics in the presence of applied currents.  We note here that the relevant quantity for vortices is the combination of the chirality and the index giving the spin direction in the vortex core.

The new contributions of the present paper include a derivation of the Landau-Lifshitz-Gilbert equations from microscopics which is based on a theory of the rotational degrees of freedom, complementing previous derivations.~\cite{Bazaliy98,Duine07,MacDonald09} We have also included the magnetic dipole energy, which is found to make a numerically large contribution. The fact that the dipole interaction couples spin and space directions leads to the dependence of critical current on angle between current and magnetization.  We have   distinguished the transport and spin relaxation scattering processes, derived an expression (following Ref.~\onlinecite{Mitra08}) for the current-induced quasithermal noise, and explicated in detail the correspondence between superconducting and magnetic vortices. The most important difference is that in the magnetic case dissipative effects are in general expected to be weak so the vortex  Hall angle is generically close to $\pi/2$.

An important question concerns the experimental observability of the transition we discuss. Current-induced  vortex motion  has been observed.~\cite{Klaui2006,Niu09,Heyn10,Pfleiderer2010} The relative magnitude of the adiabatic and non-adiabatic spin transfer torques has been determined by studying displacement of vortices in permalloy discs.~\cite{Heyn10} In Ref~\onlinecite{Klaui2006} the nucleation and annihilation of magnetic vortices due to current induced spin transfer torque, an effect which has been theoretically predicted,~\cite{Thiaville05,Shibata05} was observed in NiFe wires. Jonietz {\it et al} used neutron scattering to detect a current-induced shift in the skyrmion lattice in the helimagnet $MnSi$. It was theoretically shown in Ref~\onlinecite{Berger86}, and later experimentally confirmed~\cite{Niu09} that moving vortices can produce a ``ferro''-Josephson effect where a voltage  drop can be induced in the direction transverse to the motion of a vortex. The current induced vortex
binding unbinding transition discussed in this paper should create a similar additional voltage drop arising from the motion of free vortices. Thus the effects we predict are in principle observable.

A crucial simplifying assumption made in this paper was that the vortex-induced length scales are long enough that we may assume that the quasi-particle distribution relaxes to a steady state determined by the local magnetization. It is quite likely that in many cases this locality assumption is violated, and that issues arise analogous to those arising in the physics of vortex motion in clean superconductors. Investigating this issue is an important direction for future research. Further important open questions concern the nucleation of vortices at sample boundaries, magnetic domain walls, and near local inhomogeneities, and also the extension of our work to other cases such as monopoles in three dimensional magnets.

{\it Acknowledgement:} We thank A. Kent for helpful discussions. This work was supported by
NSF-DMR (Grant Nos. 1004589 (AM) and 1006282 (AJM)).


\end{document}